\documentclass[aps,pra,twocolumn,showpacs,preprintnumbers,amsmath,amssymb]{revtex4-1}

\usepackage{graphicx}
\usepackage{dcolumn}
\usepackage{bm}
\usepackage{natbib}
\setcitestyle{square}

\usepackage{amsmath}
\usepackage{braket}
\usepackage{mathrsfs}
\usepackage{amsfonts}
\usepackage{color}
\usepackage{physics}

\usepackage[normalem]{ulem} 

\usepackage{hyperref}
\usepackage{cleveref}

\begin{document}


\title{Inconsistency between Linearized Thomas-Fermi Approximation and Electron-Ionized Impurity Scattering Rate in the first Born Approximation}

\author{Gionni Marchetti}
\email{gionni.marchetti@kbfi.ee}

\affiliation{%
National Institute of Chemical Physics and Biophysics, R\"{a}vala 10, 15042 Tallinn, Estonia\\
}%




\date{\today}

\begin{abstract}

We show that by computing the electron-impurity scattering rate at the first order via Fermi's golden rule, and assuming that the localized impurity potential is of Yukawa form, one obtains a wave vector transfer distribution which is inconsistent with the finite temperature linearized Thomas-Fermi approximation for {\it n}-type semiconductors.  Our previous findings show that this is not the case for the carrier nondegenerate dynamics, because the average  wave vector transferred being in general negligible in this regime. Moreover, we examine the behavior of the electron-impurity differential cross-sections in the first Born approximation for relevant values of the wave vector transfer. We find that in the majority of collisions, the scattering probabilities  differ at the most by $1$ \% from the estimates computed by means of the impurity potential at random phase approximation level. 
\end{abstract}

\maketitle

\section{Introduction} \label{intro}

Point defects are ubiquitous in real crystals. Imperfections, vacancies and interstitials, can alter major physical properties of solids such as the phonon spectrum, as well as optical and electrical properties to some extent \cite{ashcroft1976}. 

Defects in semiconductors such as Si, Ge and compound semiconductors such as GaAs in general dissolve on lattice sites creating  shallow point-like acceptors or donors, which make them suitable for electronic and spintronic device applications \cite{bracht2015, zutic2004}. In general ionized impurities affect the carrier dynamics in bulk semiconductors at high doping concentrations or at low temperatures, see Ref.~ \cite{chattopadhyay1981,marchetti2017}.

Different theoretical approaches  have been proposed  to tackle the problem of interactions between point-like ionized impurities and electrons in solids. The standard tools are: the Green's function formalism \cite{doniach1974}, density functional theory (DFT) \cite{alkauskas2016,Coutinho2015,hohenberg1964} and first order time-dependent perturbation theory of quantum mechanics \cite{Weinberg2013}. Assuming  that the  impurities, which are randomly distributed centers in bulk semiconductors, cause  small perturbations, and neglecting exchange and correlation effects, one can apply the  linear response theory (LRT) in random phase approximation (RPA).  Then in the limit of small wave vector transfer, i.e. when  a  linearized Thomas-Fermi approximation (LTFA) holds, a screened Coulomb potential of Yukawa form is sufficient to accurately account for  electron-impurity (e-i) interaction. By means of this impurity potential, the scattering probability for a single (incoherent) electron-impurity collision,  is computed via the Fermi golden rule (FGR).  This important result is often  referred to as Brooks-Herring (B-H) approach \cite{brooks1951, fischetti2016,chattopadhyay1981}. Note that although this simple model neglects the perturbing effects of the impurities on the carrier energy levels and wave functions \cite{moore1967}. However, despite its simplicity, it is able of giving quantitatively accurate carrier mobilities  for most of cases of interest, and is routinely employed in ensemble Monte Carlo (EMC) simulations of carrier transport \cite{jacoboni1983,jacoboni1989}. 

The Brooks-Herring model has the amenable property to make the electron-impurity scattering problem tractable \cite{canright1988}. However, Fermi's golden rule entails the first Born approximation (B1),  whose validity becomes questionable in the low energy limit \cite{ziman79}. As a consequence, it was found that in general, the B-H approach overestimates the electron-impurity interactions at low  energies, see Ref.~\onlinecite{ridley2013}. In the following we make no attempt  to investigate the limits of the Born approximation, which have been addressed in many other papers \cite{blatt1957,kubo1965, moore1967, chattopadhyay1981}, by assuming that the electron-impurity potential is of Yukawa form. 
As the B-H model rests on the combination of two basic approximations, i.e. B1 and RPA, 
 different authors have tried to improve it  by bypassing these central approximations. One possible attempt to overcome B1, needs to include quantum corrections such as the second (B2) and third Born approximations, coherent scattering from pairs of distinct impurity centers, and dressing effects of impurities on  the carriers' energy spectrum and wave functions \cite{moore1967}. In this case, the carrier and impurities continue to interact via a Yukawa potential, i.e. the impurity potential is still evaluated at RPA level. In general, it is found that for for a  {\it n}-type  GaAs, these corrections become questionable for concentrations smaller than $10^{18}$ $\mathrm{cm}^{-3}$ \cite{moore1967,moore1967a,chattopadhyay1981}. The alternative approach to bypass the B1,  can be pursued by including the contribution to the scattering probability  from all the terms of the Born series via the  the partial-wave analysis \cite{meyer1981}.  Very recently, this was also achieved  through the phase variable method (VPM), which computes the scattering phase shifts accurately \cite{marchetti2018}. 
 
On the other hand, improving over the  RPA, by including exchange and correlations effects, is also viable \cite{giuliani2005}. In semiconductor physics literature, accounting for these many-body effects, is often referred to as Takimoto  screening \cite{takimoto1959, ridley2013}. Note that introducing these many-body corrections to the simple RPA, the impurity potential typically takes the form of  an exponent cosine screened Coulomb potential. The latter   is routinely employed in plasma physics, see Ref.~\onlinecite{eliasson2016}. Instead, in the bulk semiconductor physics,  this potential is scarcely used, as it can model the e-i interactions meaningfully only  for dopant concentrations much smaller than  $10^{17}$ $\mathrm{cm}^{-3}$ \cite{ridley2013}. 
 
In this paper, we  investigate how the Brooks-Herring model invalidates the LTFA through the electron-impurity scattering kinematics,  in the degenerate regime, i.e. for  temperatures $T$ much  smaller than Fermi temperature $T_{\mathrm{F}}$, in {\it n}-type semiconductors. To this end, we consider the carrier dynamics in bulk {\it n}-type GaAs semiconductor whose material parameters (doping, temperatures) are chosen  to ensure that the basic conditions on which the Brooks-Herring model rests, are satisfied. 

To address this inconsistency, we  study the wave vector, or equivalently the momentum, transfer due to the e-i intravalley  elastic collisions within the linear response theory for an electron liquid. The polar scattering angles, consistent with the B-H model, are stochastically generated by means of a particular EMC algorithm, for a large number of scattering events \cite{jacoboni1983,jacoboni1989}. This algorithm allows us to obtain the wave vector transfer distribution for the material parameters under scrutiny.  

Our findings show that the average wave vector transfer is   $\bar{q} \sim 0.7  k_\mathrm{F}$, where $k_\mathrm{F}$ is the Fermi wave vector. This result manifestly invalidates the linear Thomas-Fermi approximation, which  hold only for $q \simeq 0$ (or equivalently $q \ll k_\mathrm{F}$) \cite{ashcroft1976,giuliani2005}. Therefore,  the B-H approach proves  unable to prevent  large wave vector transfer. On the other hand, we previously found that in general this does not occur when  the carrier dynamics is studied in a nondegenerate regime \cite{marchetti2018}. In the latter case, the self-consistency between the LFTA and B-H model is correctly achieved.

Additionally, as the linear TF screening, does not distinguish between attractive and repulsive Coulomb interaction, we address the effects of the inverse screening  length, computed up to the second Born approximation, on the momentum transfer. It could be believed that that this screening modification, which rests on the Friedel sum rule (FSR), could prevent large momentum transfer, previously observed in the electron-impurity scattering processes. However, our findings prove that indeed it is not always ($\bar{q} \sim 0.8  k_\mathrm{F}$).

Finally, our analysis of the $q$-dependence of the B1 electron-impurity differential cross-sections, shows that small discrepancies, of about $1$ \%, arise between the scattering probabilities due to the impurity potentials in LTFA, i.e. of Yukawa form, and in RPA respectively, for the most relevant $q$ values.

This paper is organized as follows. In Section~ \ref{sec:1}  we recall the main results about the linear response theory for the electron liquid, including the derivation of the LTFA inverse screening length at finite temperature, and the derivation of the electron-impurity scattering rate in B1.  In  Section \ref{sec:2a} we 
study the wave vector transfer distribution  in {\it n}-type GaAs.  Moreover, we show that all the basic criteria for the validity of the B-H model, are completely satisfied for the chosen material parameters. In Section \ref{sec:2a} we present the effects of the B2 inverse screening on the momentum transfer.  We also include all the necessary formalism to make the screening parameter (impurity) charge-dependent. Finally, in Section \ref{sec:3} the behavior of the e-i differential cross-sections against the wave vector transfer is illustrated for the material parameters under scrutiny.  We also discuss the physical significance attached to the wave vector transfer, which mainly motivated this work.

\section{Electron-Test Charge Interaction in Linear Response Theory at RPA level}\label{sec:1}

The model of impurity screening we employ idealizes the actual situation of a realistic positive ion  embedded in a weakly interacting electron gas in a paramagnetic state.  
In fact an impurity, like proton in the Hydrogen atom, is a source of a strong perturbation to the electron gas surrounding it, giving rise to non-linearity effects stronger in the vicinity of the impurity where the electron density becomes large enough to make the LRT approach questionable \cite{jena1978,simion2005}.
In the following we will limit ourselves to the linear response theory 
In order to estimate the screened electron-impurity interaction at RPA level, we need the finite temperature density-density response function  $\chi_{nn}\left(q,\omega, T\right)$, where q, $\omega$ are the wave vector and the frequency  respectively,  for a homogeneous three-dimensional non-interacting electron gas in a paramagnetic state. The real and immaginary parts of $\chi_{nn}$ read  \cite{giuliani2005,maldague1978} 
\begin{widetext}
\begin{equation}\label{eq:lindhardReal}
 \frac{\Re  \chi_{nn}\left(q,\omega, T\right)}{N(0)}= -\int_{0}^{\infty} d\,x \frac{F\left(x,T \right)}{2\bar{ q}}
 \left(\ln \left|\frac{x-v_{-}}{x+v_{+}}\right|- \ln \left|\frac{x-v_{+}}{x+v_{+}}\right| \right)\, ,
\end{equation}
\end{widetext}
and 
\begin{equation}\label{eq:lindhardIm}
 \frac{\Im \chi_{nn}\left(q,\omega, T\right)}{N(0)}=-\frac{\pi}{2} \left( \frac{\omega}{q v_{F}}+ \frac{k_{B}T}{\hbar q v_{F}}
 \ln \frac{1+e^{\beta\left[v_{-}^{2}E_{F}-\mu\right]}}{1+e^{\beta\left[v_{+}^{2}E_{F}-\mu\right]}}\right) \, ,
\end{equation}
respectively. In Eq.~\ref{eq:lindhardReal}, Eq.~\ref{eq:lindhardIm} we have introduced the following dimensionless variables $\bar{q}=q/k_{F} $, and $v_{\pm}=\omega/q v_{F}\pm q/2 k_{F} $ where $k_{F}$ and $v_{F}= \hbar k_{F}/m^{\ast}$  are the Fermi wave vector and Fermi velocity respectively, $m^{\ast}$ being the carriers' effective mass. The symbol $N(0) \equiv m^{\ast} k_{F}/\pi^{2}\hbar^{2}$ denotes the total density of states per unit volume at the Fermi energy for an electron gas in a paramagnetic state, $\hbar$ being the reduced Planck constant.

In Eq.~\ref{eq:lindhardReal} the function $F\left(x,T\right)$ is given by
\begin{equation}\label{eq:function}
 F\left(x,T\right)=\frac{x}{e^{\beta \left[x^{2}E_{F}-\mu \right]}+1} \, ,
\end{equation}
where $E_{F}$, $\mu$ are the Fermi energy and the electronic chemical potential respectively. In Eq.~\ref{eq:lindhardIm} we define $ \beta \equiv 1/k_{B}T$, $k_{B}$ being the Boltzmann constant .

The presence of impurities would modify $\chi_{nn}$, however we shall be interested only in the static response, and hence this effect can be ignored \cite{giuliani2005,mermin1970} . The next step is to include the response of an interacting electron liquid through the dynamical RPA dielectric function $\epsilon^{\mathrm{RPA}}  \left(q, \omega, T \right)$ 
\begin{equation}\label{eq:dielectric}
\epsilon ^{\mathrm{RPA}} \left(q, \omega, T \right)= 1 - v_{q} \chi_{nn} \left(q,\omega, T \right) \, ,
\end{equation}
where $e$ and $Ze$ are the magnitudes of the elementary charge and the impurity charge respectively, and $ v_{q}= - 4\pi Ze^{2}/q^{2} $ is the Fourier transform of the bare Coulomb potential. Thus the electron-impurity test charge screened interaction at RPA level is 
\begin{equation} \label{eq:screenedV}
V_{\mathrm{ei}}^{\mathrm{RPA}}\left(q, \omega \right) = \frac{v_q}{\epsilon ^{\mathrm{RPA}} \left(q, \omega, T \right)}  \, .
\end{equation}
So far the results are quite general, indeed  the nonparabolicity and other band-structure effects are expected to give corrections of second order \cite{meyer1983}, however the impurity potential of the B-H model rests on two more crucial assumptions. First, it considers only static perturbations to the electron system. Second, it assumes that these perturbations occur only in the long wavelength limit,  i.e. $q \ll k_{\mathrm{F}}$. The latter  provides the Thomas-Fermi approximation for the electron-impurity interaction potential, and shapes the impurity potential  into a potential of Yukawa form.

As we shall consider elastic scattering between impurities and electrons, the adiabatic linear response for density fluctuations at finite $q$ in the low-frequency limit, can be obtained by setting $\omega = 0$ in Eq.~ \ref{eq:dielectric}. Finally, it is possible to expand the dynamic dielectric function (the imaginary part vanishes for $\omega \to 0$, and hence $\Re  \epsilon ^{\mathrm{RPA}}= \epsilon ^{\mathrm{RPA}}$), as the following series \cite{meyer1983}
\begin{widetext}
\begin{equation}\label{eq:expansion}
 \epsilon ^{\mathrm{RPA}} \left(q , 0, T \right)  -1 	\approx \frac{ q_0^{2}}{q^{2}}\left[1 - \frac{1}{6} \left( \frac{q^{2}\hbar^{2}}{2  m^{\ast} k_{B} T } \right) \frac{\mathscr{F}_{-3/2}}{\mathscr{F}_{-1/2}} - \frac{1}{60} \left( \frac{q^{2}\hbar^{2}}{2  m^{\ast} k_{B} T } \right)^{2} \frac{\mathscr{F}_{-5/2}}{\mathscr{F}_{-1/2}} + \cdots \right] \, ,
\end{equation}
\end{widetext}
where $\mathscr{F}_{j}$ denotes Fermi integral of order $j$ \cite{blakemore1962}, and $q_{0}$, the the finite temperature LTFA inverse screening length at finite temperature, reads
\begin{equation}\label{eq:screeningQ}
q_{0}^{2} \equiv \frac{4 \pi  n_{\mathrm{e}} e^{2}}{ k_{\mathrm{B}}T}\frac{\mathscr{F}_{-1/2}(\eta)}{\mathscr{F}_{1/2}(\eta)} \, .
\end{equation}
Note that in Eq.~\ref{eq:screeningQ}   we defined the  reduced chemical potential $\eta \equiv \mu/\left(k_{\mathrm{B}}\mathrm{T}\right)$, which implies that we are measuring the electronic energy levels with respect to the energy of conduction band (CB) edge. 

Next, we can truncate the series, given by Eq.~ \ref{eq:expansion}, and retain only the first term in the small $q$ limit approximation ($q\ll k_{\mathrm{F}}$). A direct consequence is that 
\begin{equation}\label{eq:dielectric1}
\epsilon ^{\mathrm{RPA}} \left(q , 0, T \right)= 1 + q_0^{2}/q^{2}\, ,
\end{equation}
and hence that the impurity potential, see Eq.~\ref{eq:screenedV}, becomes of Thomas-Fermi form $V_{ei} ^{\mathrm{TF}}$, that is
 \begin{equation}\label{eq:scpotential}
V_{ei} ^{\mathrm{TF}}\left(q \right)= - \frac{ 4 \pi Z e^{2}}{  q^{2} + q_0^{2} }\, .
\end{equation}

Taking the Fourier transform of  Eq.~\ref{eq:screenedV}, one obtains the Yukawa potential, i.e., $V_{ei}^{TF}\left(r\right)=-\left(Z e^{2}/ r\right)\mathrm{e}^{-q_0 r }$, $r$ denoting the interparticle distance. The Hartree potential given by Eq.~\ref{eq:scpotential} is  a direct consequence of Thomas-Fermi theory \cite{fermi1927, thomas1927}.  Here we must recall that Thomas-Fermi theory, one of the simplest density functional theories \cite{solovej2016},  is indeed a very crude approximation. In fact the dielectric function  $\epsilon ^{\mathrm{RPA}}$  is singular at $k=2k_{\mathrm{F}}$, giving rise to the long-range oscillations of electronic density at large distances from the impurity  center (Friedel oscillations) \cite{giuliani2005}, while the Thomas-Fermi theory cannot explain this phenomenon, because it simply models the electron-impurity interaction as a monotonically decreasing potential.

Next, we  derive the electron-impurity scattering rate employed in the Brooks-Herring model which rests on the interaction potential given by Eq.~\ref{eq:scpotential},  and on single-site collisions.

In practical computations, it is customary to handle the scattering between a carrier of Bloch wave vector $\mathbf{k}$ and a point-like impurity as a perturbation via Fermi's golden rule. The transition rate  $w_{\mathrm{ei}}$ for a general impurity potential $V_{\mathrm{ei}}\left(r\right) $,  reads
\begin{equation}\label{eq:fgr}
w_{\mathrm{ei}}^{1} \left(k,k'\right) = \frac{ 2 \pi}{\hbar}  |\Bra{\mathbf{k'}} V_{\mathrm{ei}} \Ket{\mathbf{k}}|^{2} \delta \left(E'-E\right) \, ,
\end{equation}
where $\mathbf{k}$, $E$ and $\mathbf{k}^{\prime}$, $E'$  denote the carrier's wave vectors and energies before and after a collision respectively. Assuming that the electron-impurity scattering is elastic, the wave vector transfer is  $\mathbf{q} = \mathbf{k} -\mathbf{k'} $ with a scattering angle $\theta \in [0, \pi]$, see the cartoon in Fig.~\ref{fig:fig1}(top right). Note that in general, the conservation of crystal momentum requires that $\mathbf{k} -\mathbf{k'} = \mathbf{q} -\mathbf{G}$ where $\mathbf{G}$ is a reciprocal wave vector. In GaAs for intravalley collisions, there are no umklapp processes, thus in our case $\mathbf{G}=0$.  Therefore the wave vectors $\mathbf{k'}$  form an Ewald sphere,  and the wave vector transfer magnitude $q$ due to a collisional event is
\begin{equation}\label{eq:scatter1}
q^{2} = 4  k^{2} \sin^{2} \left(\theta/2\right)  \, . 
\end{equation}

Clearly, the matrix element of Eq.~\ref{eq:fgr} is proportional to the $q$-component Fourier transform  of potential , i.e.,  $\propto V_{ei}\left( q \right)$. This is why the Fermi's golden rule entails the first Born approximation \cite{joachain1987}.
If we assume that the linear Thomas-Fermi screening holds, that is $q \to 0$, then $ V_{ei}\left( q \right) = V_{ei}^{\mathrm{TF}}\left( q \right) $, see Eq. ~\ref{eq:scpotential}. In the latter case, if $n_i$ denotes the doping concentration,  inserting  Eq. ~\ref{eq:scpotential} in ~\ref{eq:fgr}, one gets \cite{jacoboni2010}
\begin{equation}\label{eq:fgr1}
w_{ei}\left(k,k'\right) = \frac{ 2 \pi}{\hbar} \frac{n_i \left(4 \pi Z e^{2}  \right)^{2} e^{4}}{V  \left(q^{2} + q_0^{2}  \right)^{2}} \mathcal{G} \delta \left(E'\left(\mathbf{k'}\right)-E \left(\mathbf{k}\right)\right) \,,
\end{equation}
where $V$ is the volume of the solid, and the overlap integral $ \mathcal{G}$ for transitions between band of index $n', n$ reads \cite{antoncik1963}
\begin{equation}\label{eq:overlap}
\mathcal{G}(\mathbf{k'},\mathbf{k})=\int_{V_c}  u_{n'\mathbf{k'}}^{\ast}(\mathbf{r})u_{n\mathbf{k}}(\mathbf{r}) \, d\,\mathbf{r}
\end{equation}
 $V_c$ being the unit cell volume. The symbol $u_{n\mathbf{k}}$ denotes the modulating periodic part of the Bloch functions.
Hence, once the semiconductor band-structure is known, the electron-impurity scattering rate, and the wave vector transfer distribution can be derived analytically  through Eq. ~\ref{eq:fgr1}, see the relevant discussion in Section \ref{sec:2a}. 

Note that in this derivation  we  limited ourselves to a weak-scattering regime which ensures that Fermi golden's rule does not need a modification for including the collisional broadening \cite{davies1997}.

\section{ Wave Vector Transfer: Inverse Screening Length Computed in B1} \label{sec:2a}

In this paper  we perform sample calculations for  a {\it n}-type  GaAs. We consider carrier dynamics at the bottom of the central $\Gamma$  valley in GaAs. Despite this fact, our analysis is certainly quite general and holds for any bulk semiconductor insofar as the carriers can be considering  roaming  in an ideal spherical conduction band. Hence in the effective-mass approximation Bloch electrons have a scalar effective mass $m^{\ast}$, and the density of states (DOS) can be computed analytically through the  parabolic energy dispersion $E=\hbar^{2} k^{2}/2m^{\ast} $. The overlap integral $\mathcal{G}$ is shown  to be unity  for the intravalley transitions ($n'=n$) in a spherical CB \cite{jacoboni2010}. Moreover without loss of generality, we shall consider only the case of univalent impurities ($Z =1$). 

In the following we study wave vector transfer distribution for a doped GaAs with these band-structure parameters: $\varepsilon=12.9 \, \varepsilon_{0}$, $m^{\ast}=0.067 \, m_{\mathrm{e}}$, $\varepsilon_{0}$,  $m_{\mathrm{e}}$ being the vacuum permittivity  and the electron bare mass respectively \cite{vurgaftman2001}. The doping concentration is $n_i= 5 \times 10^{17}$ $\mathrm{cm}^{-3}$, and we shall assume that the electron density $n_e = n_i$ \cite{marchetti2017} which, ignoring the crystal lattice structure, corresponds to a homogeneous electron gas (jellium model) at  $T_{\mathrm{F}} \simeq 398 $  $\mathrm{K}$  with Wigner-Seitz radius $r_s= 0.7$ ($1/n_e \equiv  \left(4\pi/3\right) \left(r_s a_{0}^{\ast} \right)^{3}$,  $ a_{0}^{\ast}$ being the effective Bohr  radius). The latter condition guarantees that RPA holds. 

Furthermore for this choice of intermediate doping  density (smaller than  $ \approx 10^{18}$ $\mathrm{cm}^{-3}$),  one can safely ignore the possibility of multiple scattering events during the carrier dynamics \cite{fischetti2016} as well as the risk of impurity potential overlapping, and thus possible violations of the Friedel sum rule (FSR) \cite{sanborn1992}. The Meyer and Bartoli's criterion \cite{meyer1984a}, see \ref{eq:mbIneq} and the relative results, seems to suggest that we chose the right material dopant concentration.

We consider e-i scattering events for the range of temperatures  $T =32$ to $77$ $\mathrm{K}$ where carrier dynamics is expected to be degenerate ($T \ll T_{\mathrm{F}} $). The degeneracy  of the carrier distribution for this doping concentration at $T= 77$ $\mathrm{K}$ is also confirmed by experiments of how carrier heating affects the Burstein shift \cite{jantsch1977}. It was found that one needs to apply electric fields of strength $200$ to $900$ $V/\mathrm{cm}$ to observe a distinct non-Fermian behavior. Then it seems reasonable to assume in good approximation that carriers scatter off impurities as wave planes of Fermi wave vector $\mathbf{k}_{\mathrm{F}}$ ($E_\mathrm{F} \simeq 34$ meV) \cite{sakaki1980} whose magnitude is $k_{\mathrm{F}}=\left(3 \pi^{2} n_e\right)^{1/3}$ \cite{mermin1970}. 

The ensemble Monte Carlo, a  numerical method routinely employed for  solving Boltzmann transport equation in semiconductors, provides a specific algorithm to stochastically  generate the carrier-impurity collisional angles   $ \theta _r$ within B-H model, which reads \cite {jacoboni1989} 

\begin {equation}\label {eq:scatter2} 
\cos \theta _r = 1 - \frac  {2\left (1-r\right )}{1 +4 \gamma r} 
\end {equation} 
where $r$ is a uniform random number between $0$ and $1$ and $\gamma = E_\mathrm{F}/\protect \mathaccentV {tilde}07E{E}$ ($\protect \mathaccentV {tilde}07E{E} \equiv \hbar ^{2} q_{0}^{2}/2m^{\ast }$). Note that this algorithm  is derived from Eq.~ \ref {eq:fgr1}, and thus requires that the band-structure  be spherical (effective-mass approximation). Furthermore, it corresponds to a normalized probability density.

The validity criterion of B1 for the Brooks-Herring model requires \cite{chattopadhyay1981} that
\begin {equation}\label{eq:B1bounds} 
4 \gamma  \gg 1 \, ,
\end {equation} 
which certainly holds in our case ($\gamma  \approx 2$). To ascertain that the inequality Eq.~\ref{eq:B1bounds} gives a reasonable result, we also computed the exact scattering phase shifts $\delta_l$  for angular momentum numbers $l=0,1$,  which arise from solving the Schr\"{o}dinger radial equation in the presence of the impurity potential $V_{ei}^{TF}$, by means of the VPM \cite{calogero1963,calogero1967,marchetti2018}. We found that their values (in radians) are $\delta_0 \approx 0.5$, 
$\delta_1 \approx 0.2$, thereby confirming that B1 is a good approximation, being all the phase shifts much smaller than $\pi/2$ \cite{marchetti2018, morse1933,bethe1957}. 

For the material parameters under scrutiny, the multi-ion interference on electron-impurity scattering, can be also excluded. In fact, the criterion for which the single-ion site picture is formally valid, needs that the dimensionless parameter $d$, which gives the average number of impurities contained in a sphere of radius $1/q_0$,  satisfies the following inequality \cite{meyer1984a}:
\begin{equation}\label{eq:mbIneq}
d \lesssim \frac{8}{\left( 1 + 64 b^{-3/2} \right)} \, ,
\end{equation}
where $b = 4 \gamma$. In our case, we found that $d \approx 0.4 $ while the right-hand side of Eq.~\ref{eq:mbIneq} is approximately $2$. So, the assumption of independent scattering is certainly satisfied.

\begin{figure}
\resizebox{0.50\textwidth}{!}{%
  \includegraphics{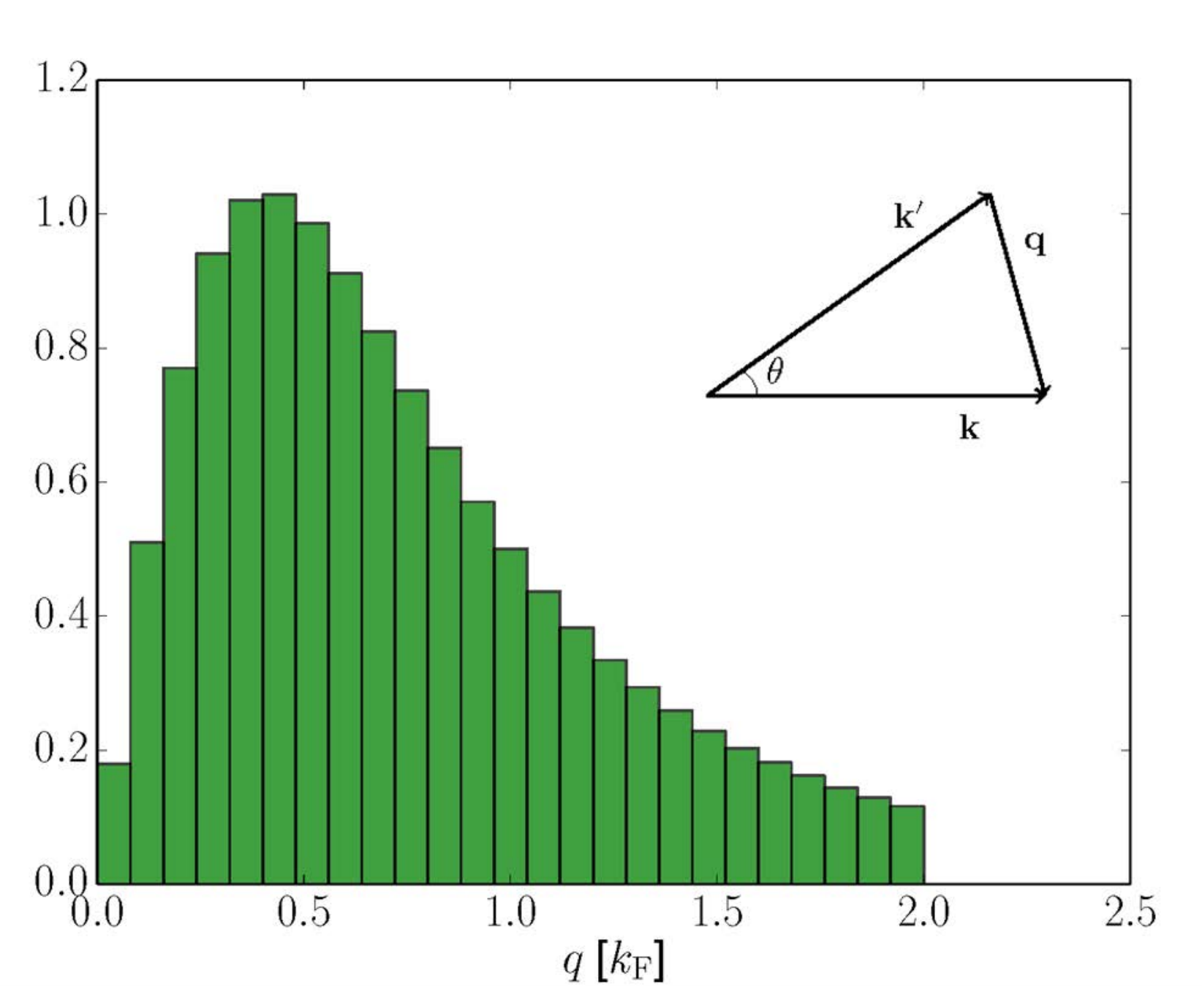}
}
\caption{Normalized histogram of wave vector transfer $q$ (in $B1$) corresponding to $10^{7}$ collisions.  In this case the average wave vector transfer $\bar{q}$ is $0.72 k_\mathrm{F}$. Here $T =32$ $\mathrm{K}$, $n_i= 5 \times 10^{17}$ $\mathrm{cm}^{-3}$. On the top right the geometrical representation of $\mathbf{q}$ after an elastic collision. Here $\mathbf{k}$ and $\mathbf{k}^{\prime}$ ($ k = k'$) are the electronic Bloch wave vectors before and after an elastic collision respectively, and $\theta$ is the (polar) scattering angle}

\label{fig:fig1}       
\end{figure}

In the following we generated $10^{7}$ e-i collisional events by means of Eq.~\ref{eq:scatter2}, for the temperatures of interest ($T =32 - 77$ $\mathrm{K}$). Note that in EMC simulations the number of collisions is usually larger, however this makes a negligible difference for present analysis. In Fig.~\ref{fig:fig1} we plot the (normalized) histogram of wave vector transfer due to electron-impurity scattering for $T = 32$ $\mathrm{K}$. In general, partial waves of different angular momentum $l$ would affect this distribution through their quantum interference. Only performing either a phase shift analysis \cite{calogero1967} or a DFT computation one can get their exact contribution, which in our case,  on the basis of reasonable semiclassical assumptions, we can infer to be s,p-waves ($l=0,1$), see Ref.~\onlinecite{marchetti2017}. Moreover from scattering theory we  expect that the partial wave $l=1$ contributes to the angular distribution for energies smaller than those significant for its contribution to the total cross-section \cite{schiff1968}. 
The distribution plotted in Fig.~\ref{fig:fig1} shows that the maximum of distribution occurs for $q_{max} \simeq 0.4k_{\mathrm{F}}$ and the wave vector transfer mean value is  $\bar{q} = 0.72k_\mathrm{F}$. Additionally, about $10\%$ of the events have large  scattering angles with $\theta \geq \pi/2$. These results clearly contradict the crucial assumption $q \ll k_\mathrm{F}$ for validity of the Thomas-Fermi approximation, and therefore  invalidate the screened Coulomb interaction given by Eq.~ \ref{eq:scpotential}. We also see that the $q^{2}$   term  in the denominator of Eq.~ \ref{eq:fgr1} is not sufficient to prevent large $q$ values. On the other hand, the study of carrier dynamics in a nondegenerate regime clearly showed that in general the wave vector  transfer $q$ during the collisions is negligible \cite{marchetti2014}. Similar results (not shown) are obtained for all the temperatures of interest.

\section{ Wave Vector Transfer: Inverse Screening Length Computed in B2} \label{sec:2b}

Next, we consider a completely different, but equivalent approach, which relates the Thomas-Fermi linear screening theory to the nonrelativistic scattering theory on the basis of charge neutrality provided by FSR. This will allow us to accurately compute the inverse screening length in the second Born approximation. The main physical reason to do so, is that the first Born approximation does not distinguish between attractive and repulsive interaction potential. Here, we need to stress that in the following we are not going to compute the scattering rate beyond the first Born approximation, i.e.,  the Fermi's golden still holds, but  the inverse screening length will be ``optimized'' to the second order in the Born series.

The Friedel sum rule (FSR) \cite{friedel1958,mahan2000} is a statement for the  complete screening of the charge impurity by the surrounding electron gas. For {\it n}-type semiconductors with one parabolic band, FSR reads \cite{stern1967}
\begin{equation}\label{eq:fsr}
 \frac{2}{\pi}\sum_{l=0} ^{\infty}\left(l+1 \right)\int_{0}^{\infty} f\left(E\right) \frac{d\delta_l\left(E\right)}{d E} d\, E= Z \, ,
\end{equation}
where  $f$ is the Fermi-Dirac distribution, and $E$ is the carrier's collisional energy. Given a general impurity potential $V_{\mathrm{ei}}$, the phase shifts can be computed in B1 by means of this formula \cite{joachain1987}
\begin{equation} \label{eq:B1shifts}
\left(\tan \delta_l \right)_{\mathrm{B1}} = - k_{\mathrm{F}} A_l
\end{equation}
with
\begin{equation}\label{eq:Al}
A_l = \int_{0}^{\infty} j_l^{2}\left(k_{\mathrm{F}}r\right) U\left(r\right) r^{2} d\, r \, ,
\end{equation}
where in Eq.~\ref{eq:Al} $j_l$ denote the spherical Bessel functions and we have introduced the reduced potential  $U \equiv \left(2 m^{\ast}V_{\mathrm{ei}}/\hbar^{2} \right) $ .
When B1 holds, then the $\delta_l$ are small, and hence $\left(\tan \delta_l \right)_{\mathrm{B1}} = \delta_l \equiv\delta_{l,\mathrm{B1}}$. In the latter case, if we insert $\delta_{l,\mathrm{B1}}$ into Eq.~\ref{eq:fsr}, we get the following mathematical constraint on the impurity potential \cite{chattopadhyay1981}
\begin{equation}\label{eq:constraint}
\frac{1}{\sqrt{\pi \beta}}\left(\frac{2 m^{\ast }}{\hbar^{2} }\right)^{1/2}\mathscr{F}_{-1/2}(\eta)\int_{0}^{\infty} U\left(r\right) r^{2} d\, r = Z\, .
\end{equation}

Now, let us assume that the impurity potential takes a Yukawa form, with unknown screening parameter $q^{\ast }$ i.e. 
\begin{equation}\label{eq:screenedCoulomb}
V_{\mathrm{ei}}\left(r\right)=-\left(Z e^{2}/ r\right)\mathrm{e}^{- q^{\ast }r } \, , 
\end{equation}
then, inserting Eq.~\ref{eq:screenedCoulomb} into Eq.~\ref{eq:constraint}, one finds that $q^{\ast } \equiv q_{0,\mathrm{B1}}=q_0$, see Eq.~ \ref{eq:screeningQ}.
This is really a remarkable result. Therefore, the inverse screening length can be obtained self-consistently requiring that the phase shifts obey the FSR. The previous procedure suggests how to compute the inverse screening length $q_{0,\mathrm{B2}}$ in second Born approximation for a potential of Yukawa form.

The agreement with the Born series through second order, can be achieved by computing $\zeta_l = \left(\tan \delta_l \right)_{\mathrm{B1}} + \left(\tan \delta_l \right)_{\mathrm{B2}}$ where the second addend defines the phase shifts in B2, see Ref. \onlinecite{joachain1987}, for its definition.

In order to avoid the calculation of the two terms separately, i.e. $\left(\tan \delta_l \right)_{\mathrm{B1}}$ and $\left(\tan \delta_l \right)_{\mathrm{B2}}$, a not easy task due to the rapidly oscillating integrand for a potential of Yukawa form, one can resort to the Schwinger variational principle for the phase shifts.
Indeed, by means of the Schwinger variational principle $\zeta_l$ is given by \cite{joachain1987,patterson1989} 
\begin{equation}\label{eq:schwinger}
\tan \zeta_l =- k_\mathrm{F} A_l \left(1-B_{l}/A_{l} \right) \, ,
\end{equation}
where 
\begin{align}\label{eq:Bl}
 B_l &= \int_{0}^{\infty} d r  \int_{0}^{\infty} d r' j_l\left(k_\mathrm{F}r\right) U\left(r\right) G_l \left(r,r'\right) \nonumber \\
 & \times U\left(r'\right) j_l\left(k_\mathrm{F}r'\right) r^{2}r'^{2} \, .
\end{align}
In Eq.~\ref{eq:Bl} we have defined the following function $
 G_l\left(r, r' \right) = k j_l \left(k r_{<} \right) \eta_l  \left(k r_{>} \right)$ 
where $\eta_l$ are the spherical Neumann functions, and $r_{<} = \min \{r,r'\}$ and $r_{>} = \max \{r,r'\}$.
In the limit of low temperatures, it is possible to approximately compute $\zeta_l$ \cite{patterson1989},  and hence by means of Eqs. ~\ref{eq:fsr} and ~\ref{eq:screenedCoulomb}, one can obtain  $q^{\ast } \equiv q_{0,B2}$ in a simple closed form
\begin{equation}\label{eq:q0B2}
 q_{0, \mathrm{B2}} = \frac{q_0^{2}}{s+\sqrt[]{s^{2}+ q_0^{2}}} \, ,
\end{equation}
where $s \equiv m^{\ast} Z e^{2}/8 \pi \epsilon \hbar^{2}$. Note that now $q_{0, \mathrm{B2}}$  depends upon the impurity charge sign through $s$,  making the sign of $Z$ discernible. Moreover, we found that for $Z=1$, $q_{0, \mathrm{B2}} > q_0$. So, the donor impurity potential range becomes shorter due to the B2 approximation.

\begin{figure}
\resizebox{0.50\textwidth}{!}{%
  \includegraphics{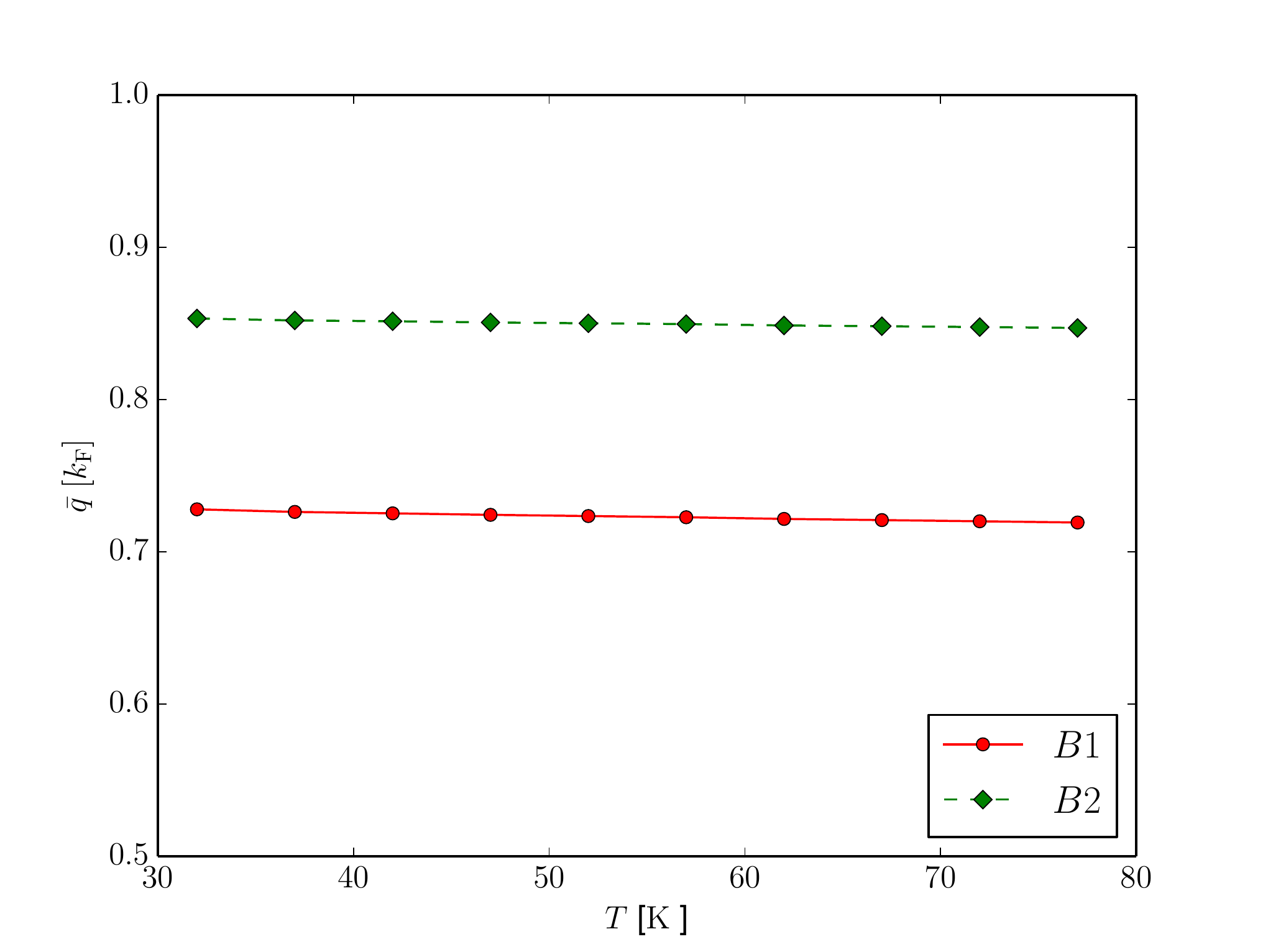}
}
\caption{Average wave vector transfer against temperatures of interest in first (B1, circles) and second (B2, diamonds)  Born approximations for Thomas-Fermi inverse screening length after $10^{7}$ collisional events ($Z=1$, $n_i= 5 \times 10^{17}$ $\mathrm{cm}^{-3}$).}
\label{fig:fig2}       
\end{figure}

The average $\bar{q}_{\mathrm{B2}}$ can be computed again via 
Eq. ~\ref{eq:scatter2} by setting $\protect \mathaccentV {tilde}07E{E} \equiv \hbar ^{2} q_{0,\mathrm{B2} }^{2}/2m^{\ast }$.
The effects on the average wave vector transfer $\bar{q}$  due to $q_{0, \mathrm{B2}}$ (diamonds) along with those of $q_0$ (circles) are illustrated in Fig.~\ref{fig:fig2} for the same number of collisions ($10^{7}$ events). 
Note that according to Ref. ~\onlinecite{marchetti2017} we expect that the B2  becomes  important for much lower carrier energies than $E_\mathrm{F}$, that is, when the scattering probability, given by Eq.~ \ref{eq:fgr1}, becomes weakly dependent on the angle $\theta$ \cite{marchetti2017}. 

In Fig.~\ref{fig:fig2},  the curves show that these averages are nearly constant: $\bar{q}_{0,\mathrm{B1}} \simeq 0.7$ and  $\bar{q}_{0,\mathrm{B2}} \simeq 0.8$ in $k_{\mathrm{F}}$ units and their variations are  about $1\%$. But  the results are even worse for B2, and are a direct consequence of the larger screening lengths due to the phase shifts computed in B2 \cite{marchetti2017}. These results confirm again that the obtained wave vector transfer distribution is  inconsistent with a linearized Thomas-Fermi approximation.

\section{Analysis of Differential Cross-Sections in Born Approximation and Concluding Remarks}\label{sec:3}

From the Fermi golden's rule is  straightforward to derive the differential cross-section $\sigma^{1} \left(\theta \right)$ in B1 for a parabolic CB. Given a general potential $V_{ei} $ one finds \cite{landau1977}
\begin{equation}
 \sigma^{1}_{ei}\left(\theta\right) = \left( \frac{m^{\ast}}{2 \pi \hbar^{2}}\right)^2 |V_{ei}\left(q \right)|^{2} \, ,
\end{equation}
where  the $\theta$-dependence is implicitly given by Eq.~\ref{eq:scatter1}. Hence, by assuming that the carriers's dynamics occur at the bottom of conduction band,  we can define the following quantity $R= \sigma_{ei}^{\mathrm{TF}}/\sigma_{ei}^{\mathrm{RPA}} $ where $\sigma_{ei}^{\mathrm{TF}}$ and $\sigma_{ei}^{\mathrm{RPA}} $  are the differential cross-sections computed in B1, using the impurity screened potentials  Eqs.~\ref{eq:scpotential}, \ref{eq:screenedV} respectively.

\begin{figure}
\resizebox{0.50\textwidth}{!}{%
  \includegraphics{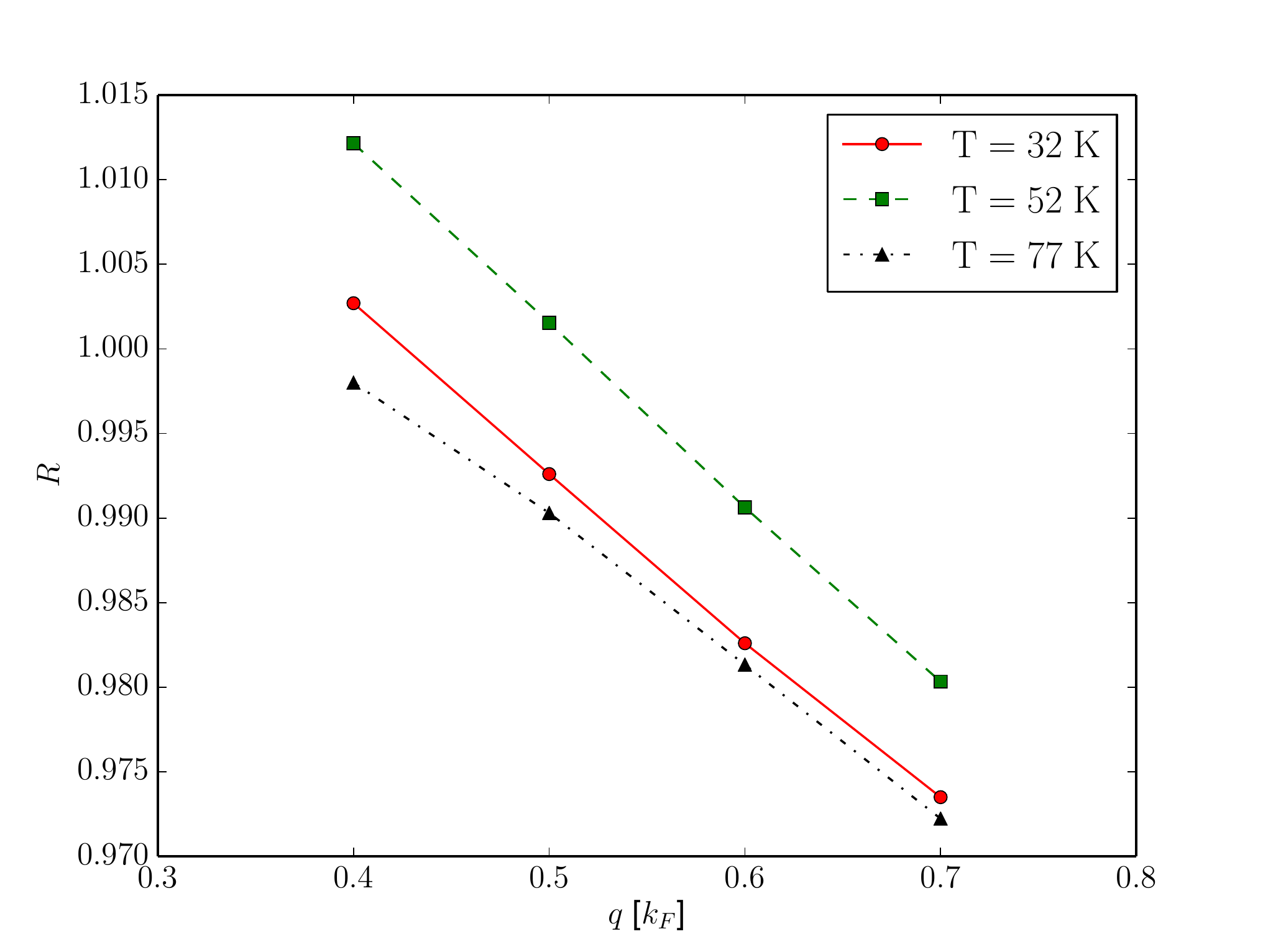}
}
\caption{ Values of $R$ against relevant values of wave vector transfer $q$ in $k_\mathrm{F}$ units for $T=32, 52, 77$ $\mathrm{K}$. The screening length $q_0$ (in B1) is computed assuming $n_i = 5 \times 10^{17}$ $cm^{-3}$.}
\label{fig:fig3}       
\end{figure}

In Fig.\ref{fig:fig3} we plot $R$ values  for some relevant wave vector transfer, in particular $q_{max}$ and $\bar{q}$ and for temperatures $T=32, 52, 77$ $\mathrm{K}$. Note that, as discussed before, the $q$ value interval of Fig.\ref{fig:fig3} is where the TF approximation completely fails. We observe that  the cross-sections differ by less than $1$ \% or  $\sim 1$ \% in the case of $T= 52$ $\mathrm{K}$ for the majority of collisions, i.e. those for which  $q=q_{max}=0.4k_{\mathrm{F}}$. For larger values of $q$ the discrepancy increases monotonically, and for  $q \to 2 k_{\mathrm{F}}$ (not shown) we found that the Brook-Herring model underestimates the scattering probability by roughly  $10$ \%. 
From this analysis we conclude that the B-H model  underestimates the scattering probabilities when they are compared with those obtained from the exact, at RPA level, impurity potential. However, by virtue of the small discrepancies at the relevant $q$ values, the B-H model still proves useful for practical applications such as EMC.

Sanborn et al. \cite{sanborn1992} argued that  strong violations of FSR (Eq.~\ref{eq:fsr}) may cause a breakdown of the linear response approximation. They stated that these violations occur, whenever the potential $V_{ei} ^{\mathrm{TF}}$ is strong enough to form its first bound state. This would mean that the first Born approximation does not longer hold. In Section \ref{sec:2a} we showed that the B1 is valid, thus ensuring that the LRT is certainly applicable to our case.

The physical significance of the momentum transfer can be easily understood from the quantum field theory, which describes the interaction of the electrons with a  charge distribution via the exchange of one, or more photons \cite{povh2002}. From this point of view,  a large momentum transfer  means that the charge distribution has a certain spatial extent. Our findings  of Sections \ref{sec:2a}, \ref{sec:2b} seem to suggest that the carriers see the impurities  as having some spatial extent for most of the collisional events, while the basic starting assumption of the theory, see Section \ref{sec:1}, is that the impurities are point-like charges. Moreover, in real semiconductors, the disorder would break the translational symmetry locally. On the other hand,  the translational symmetry underpins the linear response theory and the scattering theory, we presented in Section  \ref{sec:1}, and hence the B-H model. Thus, one may wonder how to include the effects of disorder in the present formalism, and how this may affect the wave vector transfer distribution.

\begin{acknowledgments}
We wish to express gratitude to Massimo Fischetti, Giovanni Vignale and Bengt Eliasson for some useful comments, and Marco Patriarca, Luca Marzola, Sean Fraser and David Navidad Maeso. This work was supported by IUT39-1 project of the Estonian Ministry of Education and Research.
\end{acknowledgments}

\bibliography{references}{}

\begin{thebibliography}{56}%
\makeatletter
\providecommand \@ifxundefined [1]{%
 \@ifx{#1\undefined}
}%
\providecommand \@ifnum [1]{%
 \ifnum #1\expandafter \@firstoftwo
 \else \expandafter \@secondoftwo
 \fi
}%
\providecommand \@ifx [1]{%
 \ifx #1\expandafter \@firstoftwo
 \else \expandafter \@secondoftwo
 \fi
}%
\providecommand \natexlab [1]{#1}%
\providecommand \enquote  [1]{``#1''}%
\providecommand \bibnamefont  [1]{#1}%
\providecommand \bibfnamefont [1]{#1}%
\providecommand \citenamefont [1]{#1}%
\providecommand \href@noop [0]{\@secondoftwo}%
\providecommand \href [0]{\begingroup \@sanitize@url \@href}%
\providecommand \@href[1]{\@@startlink{#1}\@@href}%
\providecommand \@@href[1]{\endgroup#1\@@endlink}%
\providecommand \@sanitize@url [0]{\catcode `\\12\catcode `\$12\catcode
  `\&12\catcode `\#12\catcode `\^12\catcode `\_12\catcode `\%12\relax}%
\providecommand \@@startlink[1]{}%
\providecommand \@@endlink[0]{}%
\providecommand \url  [0]{\begingroup\@sanitize@url \@url }%
\providecommand \@url [1]{\endgroup\@href {#1}{\urlprefix }}%
\providecommand \urlprefix  [0]{URL }%
\providecommand \Eprint [0]{\href }%
\providecommand \doibase [0]{http://dx.doi.org/}%
\providecommand \selectlanguage [0]{\@gobble}%
\providecommand \bibinfo  [0]{\@secondoftwo}%
\providecommand \bibfield  [0]{\@secondoftwo}%
\providecommand \translation [1]{[#1]}%
\providecommand \BibitemOpen [0]{}%
\providecommand \bibitemStop [0]{}%
\providecommand \bibitemNoStop [0]{.\EOS\space}%
\providecommand \EOS [0]{\spacefactor3000\relax}%
\providecommand \BibitemShut  [1]{\csname bibitem#1\endcsname}%
\let\auto@bib@innerbib\@empty
\bibitem [{\citenamefont {Ashcroft}\ and\ \citenamefont
  {Mermin}(1976)}]{ashcroft1976}%
  \BibitemOpen
  \bibfield  {author} {\bibinfo {author} {\bibfnamefont {N.~W.}\ \bibnamefont
  {Ashcroft}}\ and\ \bibinfo {author} {\bibfnamefont {M.~D.}\ \bibnamefont
  {Mermin}},\ }\href@noop {} {\emph {\bibinfo {title} {{Solid State
  Physics}}}}\ (\bibinfo  {publisher} {Saunders College},\ \bibinfo {address}
  {Philadelphia},\ \bibinfo {year} {1976})\BibitemShut {NoStop}%
\bibitem [{\citenamefont {Bracht}(2015)}]{bracht2015}%
  \BibitemOpen
  \bibfield  {author} {\bibinfo {author} {\bibfnamefont {H.}~\bibnamefont
  {Bracht}},\ }\enquote {\bibinfo {title} {Diffusion and point defects in
  silicon materials},}\ in\ \href {\doibase 10.1007/978-4-431-55800-2_1} {\emph
  {\bibinfo {booktitle} {Defects and Impurities in Silicon Materials: An
  Introduction to Atomic-Level Silicon Engineering}}},\ \bibinfo {editor}
  {edited by\ \bibinfo {editor} {\bibfnamefont {Y.}~\bibnamefont {Yoshida}}\
  and\ \bibinfo {editor} {\bibfnamefont {G.}~\bibnamefont {Langouche}}}\
  (\bibinfo  {publisher} {Springer Japan},\ \bibinfo {address} {Tokyo},\
  \bibinfo {year} {2015})\ pp.\ \bibinfo {pages} {1--67}\BibitemShut {NoStop}%
\bibitem [{\citenamefont {Zutic}\ \emph {et~al.}(2004)\citenamefont {Zutic},
  \citenamefont {Fabian},\ and\ \citenamefont {Sarma}}]{zutic2004}%
  \BibitemOpen
  \bibfield  {author} {\bibinfo {author} {\bibfnamefont {I.}~\bibnamefont
  {Zutic}}, \bibinfo {author} {\bibfnamefont {J.}~\bibnamefont {Fabian}}, \
  and\ \bibinfo {author} {\bibfnamefont {S.~D.}\ \bibnamefont {Sarma}},\
  }\href@noop {} {\bibfield  {journal} {\bibinfo  {journal} {Rev. Mod. Phys.}\
  }\textbf {\bibinfo {volume} {76}},\ \bibinfo {pages} {323} (\bibinfo {year}
  {2004})}\BibitemShut {NoStop}%
\bibitem [{\citenamefont {Chattopadhyay}\ and\ \citenamefont
  {Queisser}(1981)}]{chattopadhyay1981}%
  \BibitemOpen
  \bibfield  {author} {\bibinfo {author} {\bibfnamefont {D.}~\bibnamefont
  {Chattopadhyay}}\ and\ \bibinfo {author} {\bibfnamefont {H.~J.}\ \bibnamefont
  {Queisser}},\ }\href {\doibase 10.1103/RevModPhys.53.745} {\bibfield
  {journal} {\bibinfo  {journal} {Rev. Mod. Phys.}\ }\textbf {\bibinfo {volume}
  {53}},\ \bibinfo {pages} {745} (\bibinfo {year} {1981})}\BibitemShut
  {NoStop}%
\bibitem [{\citenamefont {Marchetti}\ and\ \citenamefont
  {D'Amico}(2017)}]{marchetti2017}%
  \BibitemOpen
  \bibfield  {author} {\bibinfo {author} {\bibfnamefont {G.}~\bibnamefont
  {Marchetti}}\ and\ \bibinfo {author} {\bibfnamefont {I.}~\bibnamefont
  {D'Amico}},\ }\href {\doibase 10.1002/pssb.201600806} {\bibfield  {journal}
  {\bibinfo  {journal} {Physica Status Solidi (b)}\ }\textbf {\bibinfo {volume}
  {254}},\ \bibinfo {pages} {e201600806} (\bibinfo {year} {2017})}\BibitemShut
  {NoStop}%
\bibitem [{\citenamefont {Doniach}\ and\ \citenamefont
  {Sondheimer}(1974)}]{doniach1974}%
  \BibitemOpen
  \bibfield  {author} {\bibinfo {author} {\bibfnamefont {S.}~\bibnamefont
  {Doniach}}\ and\ \bibinfo {author} {\bibfnamefont {E.~H.}\ \bibnamefont
  {Sondheimer}},\ }\href@noop {} {\emph {\bibinfo {title} {{Green's Functions
  for Solid State Physicists }}}}\ (\bibinfo  {publisher} {Benjamin},\ \bibinfo
  {address} {Reading, MA},\ \bibinfo {year} {1974})\BibitemShut {NoStop}%
\bibitem [{\citenamefont {Alkauskas}\ \emph {et~al.}(2016)\citenamefont
  {Alkauskas}, \citenamefont {McCluskey},\ and\ \citenamefont {Van~de
  Walle}}]{alkauskas2016}%
  \BibitemOpen
  \bibfield  {author} {\bibinfo {author} {\bibfnamefont {A.}~\bibnamefont
  {Alkauskas}}, \bibinfo {author} {\bibfnamefont {M.~D.}\ \bibnamefont
  {McCluskey}}, \ and\ \bibinfo {author} {\bibfnamefont {C.~G.}\ \bibnamefont
  {Van~de Walle}},\ }\href {\doibase 10.1063/1.4948245} {\bibfield  {journal}
  {\bibinfo  {journal} {Journal of Applied Physics}\ }\textbf {\bibinfo
  {volume} {119}},\ \bibinfo {pages} {181101} (\bibinfo {year}
  {2016})}\BibitemShut {NoStop}%
\bibitem [{\citenamefont {Coutinho}(2015)}]{Coutinho2015}%
  \BibitemOpen
  \bibfield  {author} {\bibinfo {author} {\bibfnamefont {J.}~\bibnamefont
  {Coutinho}},\ }\enquote {\bibinfo {title} {Density functional modeling of
  defects and impurities in silicon materials},}\ in\ \href {\doibase
  10.1007/978-4-431-55800-2_2} {\emph {\bibinfo {booktitle} {Defects and
  Impurities in Silicon Materials: An Introduction to Atomic-Level Silicon
  Engineering}}},\ \bibinfo {editor} {edited by\ \bibinfo {editor}
  {\bibfnamefont {Y.}~\bibnamefont {Yoshida}}\ and\ \bibinfo {editor}
  {\bibfnamefont {G.}~\bibnamefont {Langouche}}}\ (\bibinfo  {publisher}
  {Springer Japan},\ \bibinfo {address} {Tokyo},\ \bibinfo {year} {2015})\ pp.\
  \bibinfo {pages} {69--127}\BibitemShut {NoStop}%
\bibitem [{\citenamefont {Hohenberg}\ and\ \citenamefont
  {Kohn}(1964)}]{hohenberg1964}%
  \BibitemOpen
  \bibfield  {author} {\bibinfo {author} {\bibfnamefont {P.}~\bibnamefont
  {Hohenberg}}\ and\ \bibinfo {author} {\bibfnamefont {W.}~\bibnamefont
  {Kohn}},\ }\href {\doibase 10.1103/PhysRev.136.B864} {\bibfield  {journal}
  {\bibinfo  {journal} {Phys. Rev.}\ }\textbf {\bibinfo {volume} {136}},\
  \bibinfo {pages} {B864} (\bibinfo {year} {1964})}\BibitemShut {NoStop}%
\bibitem [{\citenamefont {Weinberg}(2013)}]{Weinberg2013}%
  \BibitemOpen
  \bibfield  {author} {\bibinfo {author} {\bibfnamefont {S.}~\bibnamefont
  {Weinberg}},\ }\href@noop {} {\emph {\bibinfo {title} {{Lectures on quantum
  mechanics}}}}\ (\bibinfo  {publisher} {Cambridge Univ. Press},\ \bibinfo
  {address} {Cambridge},\ \bibinfo {year} {2013})\BibitemShut {NoStop}%
\bibitem [{\citenamefont {Brooks}(1951)}]{brooks1951}%
  \BibitemOpen
  \bibfield  {author} {\bibinfo {author} {\bibfnamefont {H.}~\bibnamefont
  {Brooks}},\ }\href@noop {} {\bibfield  {journal} {\bibinfo  {journal} {Phys.
  Rev}\ }\textbf {\bibinfo {volume} {83}},\ \bibinfo {pages} {879} (\bibinfo
  {year} {1951})}\BibitemShut {NoStop}%
\bibitem [{\citenamefont {Fischetti}\ and\ \citenamefont
  {Vandenberghe}(2016)}]{fischetti2016}%
  \BibitemOpen
  \bibfield  {author} {\bibinfo {author} {\bibfnamefont {M.}~\bibnamefont
  {Fischetti}}\ and\ \bibinfo {author} {\bibfnamefont {W.~G.}\ \bibnamefont
  {Vandenberghe}},\ }\href@noop {} {\emph {\bibinfo {title} {{Advanced Physics
  of Electron Transport in Semiconductors and Nanostructures}}}}\ (\bibinfo
  {publisher} {Springer International Publishing},\ \bibinfo {address}
  {Switzerland},\ \bibinfo {year} {2016})\BibitemShut {NoStop}%
\bibitem [{\citenamefont {Moore}(1967{\natexlab{a}})}]{moore1967}%
  \BibitemOpen
  \bibfield  {author} {\bibinfo {author} {\bibfnamefont {E.~J.}\ \bibnamefont
  {Moore}},\ }\href {\doibase 10.1103/PhysRev.160.607} {\bibfield  {journal}
  {\bibinfo  {journal} {Phys. Rev.}\ }\textbf {\bibinfo {volume} {160}},\
  \bibinfo {pages} {607} (\bibinfo {year} {1967}{\natexlab{a}})}\BibitemShut
  {NoStop}%
\bibitem [{\citenamefont {Jacoboni}\ and\ \citenamefont
  {Reggiani}(1983)}]{jacoboni1983}%
  \BibitemOpen
  \bibfield  {author} {\bibinfo {author} {\bibfnamefont {C.}~\bibnamefont
  {Jacoboni}}\ and\ \bibinfo {author} {\bibfnamefont {L.}~\bibnamefont
  {Reggiani}},\ }\href {\doibase 10.1103/RevModPhys.55.645} {\bibfield
  {journal} {\bibinfo  {journal} {Rev. Mod. Phys.}\ }\textbf {\bibinfo {volume}
  {55}},\ \bibinfo {pages} {645} (\bibinfo {year} {1983})}\BibitemShut
  {NoStop}%
\bibitem [{\citenamefont {Jacoboni}\ and\ \citenamefont
  {Lugli}(1989)}]{jacoboni1989}%
  \BibitemOpen
  \bibfield  {author} {\bibinfo {author} {\bibfnamefont {C.}~\bibnamefont
  {Jacoboni}}\ and\ \bibinfo {author} {\bibfnamefont {P.}~\bibnamefont
  {Lugli}},\ }\href@noop {} {\emph {\bibinfo {title} {{The Monte Carlo Method
  for Semiconductor Device Simulation}}}}\ (\bibinfo  {publisher}
  {Springer-Verlag},\ \bibinfo {address} {Wien New York},\ \bibinfo {year}
  {1989})\BibitemShut {NoStop}%
\bibitem [{\citenamefont {Canright}(1988)}]{canright1988}%
  \BibitemOpen
  \bibfield  {author} {\bibinfo {author} {\bibfnamefont {G.~S.}\ \bibnamefont
  {Canright}},\ }\href {\doibase 10.1103/PhysRevB.38.1647} {\bibfield
  {journal} {\bibinfo  {journal} {Phys. Rev. B}\ }\textbf {\bibinfo {volume}
  {38}},\ \bibinfo {pages} {1647} (\bibinfo {year} {1988})}\BibitemShut
  {NoStop}%
\bibitem [{\citenamefont {Ziman}(1979)}]{ziman79}%
  \BibitemOpen
  \bibfield  {author} {\bibinfo {author} {\bibfnamefont {J.~M.}\ \bibnamefont
  {Ziman}},\ }\href@noop {} {\emph {\bibinfo {title} {{Principles of the Theory
  of Solids}}}}\ (\bibinfo  {publisher} {Cambridge University Press},\ \bibinfo
  {address} {Cambridge},\ \bibinfo {year} {1979})\BibitemShut {NoStop}%
\bibitem [{\citenamefont {Ridley}(2013)}]{ridley2013}%
  \BibitemOpen
  \bibfield  {author} {\bibinfo {author} {\bibfnamefont {B.~K.}\ \bibnamefont
  {Ridley}},\ }\href@noop {} {\emph {\bibinfo {title} {{Quantum Processes in
  Semiconductors}}}},\ \bibinfo {edition} {5th}\ ed.\ (\bibinfo  {publisher}
  {Oxford University Press},\ \bibinfo {address} {Oxford},\ \bibinfo {year}
  {2013})\BibitemShut {NoStop}%
\bibitem [{\citenamefont {Blatt}(1957)}]{blatt1957}%
  \BibitemOpen
  \bibfield  {author} {\bibinfo {author} {\bibfnamefont {F.~J.}\ \bibnamefont
  {Blatt}},\ }\href@noop {} {\bibfield  {journal} {\bibinfo  {journal} {J.
  Phys. Chem. Solids}\ }\textbf {\bibinfo {volume} {1}} (\bibinfo {year}
  {1957})}\BibitemShut {NoStop}%
\bibitem [{\citenamefont {Kubo}\ \emph {et~al.}(1965)\citenamefont {Kubo},
  \citenamefont {J},\ and\ \citenamefont {N.}}]{kubo1965}%
  \BibitemOpen
  \bibfield  {author} {\bibinfo {author} {\bibfnamefont {R.}~\bibnamefont
  {Kubo}}, \bibinfo {author} {\bibfnamefont {M.~S.}\ \bibnamefont {J}}, \ and\
  \bibinfo {author} {\bibfnamefont {H.}~\bibnamefont {N.}},\ }\href@noop {}
  {\emph {\bibinfo {title} {{Solid State Physics, ed. F. Seitz and D. Tumbull
  }}}}\ (\bibinfo  {publisher} {Academic Press},\ \bibinfo {address} {New
  York},\ \bibinfo {year} {1965})\BibitemShut {NoStop}%
\bibitem [{\citenamefont {Moore}(1967{\natexlab{b}})}]{moore1967a}%
  \BibitemOpen
  \bibfield  {author} {\bibinfo {author} {\bibfnamefont {E.~J.}\ \bibnamefont
  {Moore}},\ }\href {\doibase 10.1103/PhysRev.160.618} {\bibfield  {journal}
  {\bibinfo  {journal} {Phys. Rev.}\ }\textbf {\bibinfo {volume} {160}},\
  \bibinfo {pages} {618} (\bibinfo {year} {1967}{\natexlab{b}})}\BibitemShut
  {NoStop}%
\bibitem [{\citenamefont {Meyer}\ and\ \citenamefont
  {Bartoli}(1981)}]{meyer1981}%
  \BibitemOpen
  \bibfield  {author} {\bibinfo {author} {\bibfnamefont {J.~R.}\ \bibnamefont
  {Meyer}}\ and\ \bibinfo {author} {\bibfnamefont {F.~J.}\ \bibnamefont
  {Bartoli}},\ }\href {\doibase 10.1103/PhysRevB.23.5413} {\bibfield  {journal}
  {\bibinfo  {journal} {Phys. Rev. B}\ }\textbf {\bibinfo {volume} {23}},\
  \bibinfo {pages} {5413} (\bibinfo {year} {1981})}\BibitemShut {NoStop}%
\bibitem [{\citenamefont {{Marchetti}}(2018)}]{marchetti2018}%
  \BibitemOpen
  \bibfield  {author} {\bibinfo {author} {\bibfnamefont {G.}~\bibnamefont
  {{Marchetti}}},\ }\href@noop {} {\bibfield  {journal} {\bibinfo  {journal}
  {ArXiv e-prints}\ } (\bibinfo {year} {2018})},\ \Eprint
  {http://arxiv.org/abs/1804.11337} {arXiv:1804.11337 [cond-mat.mtrl-sci]}
  \BibitemShut {NoStop}%
\bibitem [{\citenamefont {Giuliani}\ and\ \citenamefont
  {Vignale}(2005)}]{giuliani2005}%
  \BibitemOpen
  \bibfield  {author} {\bibinfo {author} {\bibfnamefont {G.}~\bibnamefont
  {Giuliani}}\ and\ \bibinfo {author} {\bibfnamefont {G.}~\bibnamefont
  {Vignale}},\ }\href@noop {} {\emph {\bibinfo {title} {{Quantum Theory of
  Electron Liquid}}}}\ (\bibinfo  {publisher} {Cambridge University Press},\
  \bibinfo {address} {Cambridge},\ \bibinfo {year} {2005})\BibitemShut
  {NoStop}%
\bibitem [{\citenamefont {Takimoto}(1959)}]{takimoto1959}%
  \BibitemOpen
  \bibfield  {author} {\bibinfo {author} {\bibfnamefont {N.}~\bibnamefont
  {Takimoto}},\ }\href@noop {} {\bibfield  {journal} {\bibinfo  {journal} {J.
  Phys. Soc. Jpn}\ }\textbf {\bibinfo {volume} {14}} (\bibinfo {year}
  {1959})}\BibitemShut {NoStop}%
\bibitem [{\citenamefont {Eliasson}\ and\ \citenamefont
  {Akbari-Moghanjoughi}(2016)}]{eliasson2016}%
  \BibitemOpen
  \bibfield  {author} {\bibinfo {author} {\bibfnamefont {B.}~\bibnamefont
  {Eliasson}}\ and\ \bibinfo {author} {\bibfnamefont {M.}~\bibnamefont
  {Akbari-Moghanjoughi}},\ }\href {\doibase
  https://doi.org/10.1016/j.physleta.2016.05.043} {\bibfield  {journal}
  {\bibinfo  {journal} {Physics Letters A}\ }\textbf {\bibinfo {volume}
  {380}},\ \bibinfo {pages} {2518 } (\bibinfo {year} {2016})}\BibitemShut
  {NoStop}%
\bibitem [{\citenamefont {Jena}\ and\ \citenamefont {Singwi}(1978)}]{jena1978}%
  \BibitemOpen
  \bibfield  {author} {\bibinfo {author} {\bibfnamefont {P.}~\bibnamefont
  {Jena}}\ and\ \bibinfo {author} {\bibfnamefont {K.~S.}\ \bibnamefont
  {Singwi}},\ }\href {\doibase 10.1103/PhysRevB.17.3518} {\bibfield  {journal}
  {\bibinfo  {journal} {Phys. Rev. B}\ }\textbf {\bibinfo {volume} {17}},\
  \bibinfo {pages} {3518} (\bibinfo {year} {1978})}\BibitemShut {NoStop}%
\bibitem [{\citenamefont {Simion}\ and\ \citenamefont
  {Giuliani}(2005)}]{simion2005}%
  \BibitemOpen
  \bibfield  {author} {\bibinfo {author} {\bibfnamefont {G.~E.}\ \bibnamefont
  {Simion}}\ and\ \bibinfo {author} {\bibfnamefont {G.~F.}\ \bibnamefont
  {Giuliani}},\ }\href {\doibase 10.1103/PhysRevB.72.045127} {\bibfield
  {journal} {\bibinfo  {journal} {Phys. Rev. B}\ }\textbf {\bibinfo {volume}
  {72}},\ \bibinfo {pages} {045127} (\bibinfo {year} {2005})}\BibitemShut
  {NoStop}%
\bibitem [{\citenamefont {Maldague}(1978)}]{maldague1978}%
  \BibitemOpen
  \bibfield  {author} {\bibinfo {author} {\bibfnamefont {P.~F.}\ \bibnamefont
  {Maldague}},\ }\href@noop {} {\bibfield  {journal} {\bibinfo  {journal}
  {Surface Science}\ }\textbf {\bibinfo {volume} {73}},\ \bibinfo {pages} {296}
  (\bibinfo {year} {1978})}\BibitemShut {NoStop}%
\bibitem [{\citenamefont {Mermin}(1970)}]{mermin1970}%
  \BibitemOpen
  \bibfield  {author} {\bibinfo {author} {\bibfnamefont {N.~D.}\ \bibnamefont
  {Mermin}},\ }\href {\doibase 10.1103/PhysRevB.1.2362} {\bibfield  {journal}
  {\bibinfo  {journal} {Phys. Rev. B}\ }\textbf {\bibinfo {volume} {1}},\
  \bibinfo {pages} {2362} (\bibinfo {year} {1970})}\BibitemShut {NoStop}%
\bibitem [{\citenamefont {Meyer}\ and\ \citenamefont
  {Bartoli}(1983)}]{meyer1983}%
  \BibitemOpen
  \bibfield  {author} {\bibinfo {author} {\bibfnamefont {J.~R.}\ \bibnamefont
  {Meyer}}\ and\ \bibinfo {author} {\bibfnamefont {F.~J.}\ \bibnamefont
  {Bartoli}},\ }\href@noop {} {\bibfield  {journal} {\bibinfo  {journal} {Phys.
  Rev. B}\ }\textbf {\bibinfo {volume} {28}},\ \bibinfo {pages} {915} (\bibinfo
  {year} {1983})}\BibitemShut {NoStop}%
\bibitem [{\citenamefont {Blakemore}(1962)}]{blakemore1962}%
  \BibitemOpen
  \bibfield  {author} {\bibinfo {author} {\bibfnamefont {J.~S.}\ \bibnamefont
  {Blakemore}},\ }\href@noop {} {\emph {\bibinfo {title} {{Semiconductor
  Statistics}}}}\ (\bibinfo  {publisher} {Pergamon},\ \bibinfo {year}
  {1962})\BibitemShut {NoStop}%
\bibitem [{\citenamefont {Fermi}(1927)}]{fermi1927}%
  \BibitemOpen
  \bibfield  {author} {\bibinfo {author} {\bibfnamefont {E.}~\bibnamefont
  {Fermi}},\ }\href@noop {} {\bibfield  {journal} {\bibinfo  {journal} {Rend.
  Accad. Nat. Lincei.}\ }\textbf {\bibinfo {volume} {6}},\ \bibinfo {pages}
  {602} (\bibinfo {year} {1927})}\BibitemShut {NoStop}%
\bibitem [{\citenamefont {Thomas}(1927)}]{thomas1927}%
  \BibitemOpen
  \bibfield  {author} {\bibinfo {author} {\bibfnamefont {L.~H.}\ \bibnamefont
  {Thomas}},\ }\href@noop {} {\bibfield  {journal} {\bibinfo  {journal} {Proc.
  Cambridge. Philos. Soc.}\ }\textbf {\bibinfo {volume} {23}},\ \bibinfo
  {pages} {542} (\bibinfo {year} {1927})}\BibitemShut {NoStop}%
\bibitem [{\citenamefont {Solovej}(2016)}]{solovej2016}%
  \BibitemOpen
  \bibfield  {author} {\bibinfo {author} {\bibfnamefont {J.~P.}\ \bibnamefont
  {Solovej}},\ }\href {\doibase 10.1080/00268976.2015.1130273} {\bibfield
  {journal} {\bibinfo  {journal} {Molecular Physics}\ }\textbf {\bibinfo
  {volume} {114}},\ \bibinfo {pages} {1036} (\bibinfo {year} {2016})},\ \Eprint
  {http://arxiv.org/abs/https://doi.org/10.1080/00268976.2015.1130273}
  {https://doi.org/10.1080/00268976.2015.1130273} \BibitemShut {NoStop}%
\bibitem [{\citenamefont {Joachain}(1987)}]{joachain1987}%
  \BibitemOpen
  \bibfield  {author} {\bibinfo {author} {\bibfnamefont {C.~J.}\ \bibnamefont
  {Joachain}},\ }\href@noop {} {\emph {\bibinfo {title} {{Quantum Collision
  Theory}}}}\ (\bibinfo  {publisher} {North-Holland Physics Publishing},\
  \bibinfo {address} {Amsterdam},\ \bibinfo {year} {1987})\BibitemShut
  {NoStop}%
\bibitem [{\citenamefont {Jacoboni}(2010)}]{jacoboni2010}%
  \BibitemOpen
  \bibfield  {author} {\bibinfo {author} {\bibfnamefont {C.}~\bibnamefont
  {Jacoboni}},\ }\href@noop {} {\emph {\bibinfo {title} {{Theory of Electron
  Transport in Semiconductors}}}}\ (\bibinfo  {publisher} {Springer-Verlag},\
  \bibinfo {address} {Berlin Heidelberg},\ \bibinfo {year} {2010})\BibitemShut
  {NoStop}%
\bibitem [{\citenamefont {Antoncik}\ and\ \citenamefont
  {Landsberg}(1963)}]{antoncik1963}%
  \BibitemOpen
  \bibfield  {author} {\bibinfo {author} {\bibfnamefont {A.}~\bibnamefont
  {Antoncik}}\ and\ \bibinfo {author} {\bibfnamefont {P.~T.}\ \bibnamefont
  {Landsberg}},\ }\href@noop {} {\bibfield  {journal} {\bibinfo  {journal}
  {Proc. Phys. Soc}\ }\textbf {\bibinfo {volume} {83}} (\bibinfo {year}
  {1963})}\BibitemShut {NoStop}%
\bibitem [{\citenamefont {Davies}(1997)}]{davies1997}%
  \BibitemOpen
  \bibfield  {author} {\bibinfo {author} {\bibfnamefont {J.~H.}\ \bibnamefont
  {Davies}},\ }\href@noop {} {\emph {\bibinfo {title} {{The Physics of
  Low-dimensional Semiconductors}}}}\ (\bibinfo  {publisher} {Cambridge
  University Press},\ \bibinfo {address} {Cambridge},\ \bibinfo {year}
  {1997})\BibitemShut {NoStop}%
\bibitem [{\citenamefont {Vurgaftman}\ \emph {et~al.}(2001)\citenamefont
  {Vurgaftman}, \citenamefont {Meyer},\ and\ \citenamefont
  {Ram-Mohan}}]{vurgaftman2001}%
  \BibitemOpen
  \bibfield  {author} {\bibinfo {author} {\bibfnamefont {I.}~\bibnamefont
  {Vurgaftman}}, \bibinfo {author} {\bibfnamefont {J.~R.}\ \bibnamefont
  {Meyer}}, \ and\ \bibinfo {author} {\bibfnamefont {L.~R.}\ \bibnamefont
  {Ram-Mohan}},\ }\href@noop {} {\bibfield  {journal} {\bibinfo  {journal}
  {Journal of Applied Physics}\ }\textbf {\bibinfo {volume} {89}},\ \bibinfo
  {pages} {5815} (\bibinfo {year} {2001})}\BibitemShut {NoStop}%
\bibitem [{\citenamefont {Sanborn}\ \emph {et~al.}(1992)\citenamefont
  {Sanborn}, \citenamefont {Allen},\ and\ \citenamefont {Mahan}}]{sanborn1992}%
  \BibitemOpen
  \bibfield  {author} {\bibinfo {author} {\bibfnamefont {B.~A.}\ \bibnamefont
  {Sanborn}}, \bibinfo {author} {\bibfnamefont {P.~B.}\ \bibnamefont {Allen}},
  \ and\ \bibinfo {author} {\bibfnamefont {G.~D.}\ \bibnamefont {Mahan}},\
  }\href {\doibase 10.1103/PhysRevB.46.15123} {\bibfield  {journal} {\bibinfo
  {journal} {Phys. Rev. B}\ }\textbf {\bibinfo {volume} {46}},\ \bibinfo
  {pages} {15123} (\bibinfo {year} {1992})}\BibitemShut {NoStop}%
\bibitem [{\citenamefont {Meyer}\ and\ \citenamefont
  {Bartoli}(1984)}]{meyer1984a}%
  \BibitemOpen
  \bibfield  {author} {\bibinfo {author} {\bibfnamefont {J.~R.}\ \bibnamefont
  {Meyer}}\ and\ \bibinfo {author} {\bibfnamefont {F.~J.}\ \bibnamefont
  {Bartoli}},\ }\href@noop {} {\bibfield  {journal} {\bibinfo  {journal} {Phys.
  Rev. B}\ }\textbf {\bibinfo {volume} {30}},\ \bibinfo {pages} {1026}
  (\bibinfo {year} {1984})}\BibitemShut {NoStop}%
\bibitem [{\citenamefont {Jantsch}\ and\ \citenamefont
  {Br\"ucker}(1977)}]{jantsch1977}%
  \BibitemOpen
  \bibfield  {author} {\bibinfo {author} {\bibfnamefont {W.}~\bibnamefont
  {Jantsch}}\ and\ \bibinfo {author} {\bibfnamefont {H.}~\bibnamefont
  {Br\"ucker}},\ }\href {\doibase 10.1103/PhysRevB.15.4014} {\bibfield
  {journal} {\bibinfo  {journal} {Phys. Rev. B}\ }\textbf {\bibinfo {volume}
  {15}},\ \bibinfo {pages} {4014} (\bibinfo {year} {1977})}\BibitemShut
  {NoStop}%
\bibitem [{\citenamefont {Sakaki}(1980)}]{sakaki1980}%
  \BibitemOpen
  \bibfield  {author} {\bibinfo {author} {\bibfnamefont {H.}~\bibnamefont
  {Sakaki}},\ }\href {http://stacks.iop.org/1347-4065/19/i=12/a=L735}
  {\bibfield  {journal} {\bibinfo  {journal} {Japanese Journal of Applied
  Physics}\ }\textbf {\bibinfo {volume} {19}},\ \bibinfo {pages} {L735}
  (\bibinfo {year} {1980})}\BibitemShut {NoStop}%
\bibitem [{\citenamefont {Calogero}(1963)}]{calogero1963}%
  \BibitemOpen
  \bibfield  {author} {\bibinfo {author} {\bibfnamefont {F.}~\bibnamefont
  {Calogero}},\ }\href@noop {} {\bibfield  {journal} {\bibinfo  {journal}
  {Nuovo Cimento}\ }\textbf {\bibinfo {volume} {27}},\ \bibinfo {pages} {745}
  (\bibinfo {year} {1963})}\BibitemShut {NoStop}%
\bibitem [{\citenamefont {Calogero}(1967)}]{calogero1967}%
  \BibitemOpen
  \bibfield  {author} {\bibinfo {author} {\bibfnamefont {F.}~\bibnamefont
  {Calogero}},\ }\href@noop {} {\emph {\bibinfo {title} {{Variable Phase
  Approach to Potential Scattering }}}}\ (\bibinfo  {publisher} {Academic
  Press},\ \bibinfo {year} {1967})\BibitemShut {NoStop}%
\bibitem [{\citenamefont {Morse}\ and\ \citenamefont
  {Allis}(1933)}]{morse1933}%
  \BibitemOpen
  \bibfield  {author} {\bibinfo {author} {\bibfnamefont {P.~M.}\ \bibnamefont
  {Morse}}\ and\ \bibinfo {author} {\bibfnamefont {W.~P.}\ \bibnamefont
  {Allis}},\ }\href@noop {} {\bibfield  {journal} {\bibinfo  {journal} {Phys.
  Rev.}\ }\textbf {\bibinfo {volume} {44}},\ \bibinfo {pages} {269} (\bibinfo
  {year} {1933})}\BibitemShut {NoStop}%
\bibitem [{\citenamefont {Bethe}\ and\ \citenamefont
  {Salpeter}(1957)}]{bethe1957}%
  \BibitemOpen
  \bibfield  {author} {\bibinfo {author} {\bibfnamefont {H.~A.}\ \bibnamefont
  {Bethe}}\ and\ \bibinfo {author} {\bibfnamefont {E.~E.}\ \bibnamefont
  {Salpeter}},\ }\href@noop {} {\emph {\bibinfo {title} {{Quantum Mechanics of
  One- and Two-Electron Atoms}}}}\ (\bibinfo  {publisher} {Springer-Verlag},\
  \bibinfo {address} {Berlin G{\"o}ttingen Heidelberg},\ \bibinfo {year}
  {1957})\BibitemShut {NoStop}%
\bibitem [{\citenamefont {Schiff}(1968)}]{schiff1968}%
  \BibitemOpen
  \bibfield  {author} {\bibinfo {author} {\bibfnamefont {L.~I.}\ \bibnamefont
  {Schiff}},\ }\href@noop {} {\emph {\bibinfo {title} {{Quantum Mechanics}}}}\
  (\bibinfo  {publisher} {McGraw-Hill},\ \bibinfo {address} {Singapore},\
  \bibinfo {year} {1968})\BibitemShut {NoStop}%
\bibitem [{\citenamefont {Marchetti}\ \emph {et~al.}(2014)\citenamefont
  {Marchetti}, \citenamefont {Hodgson}, \citenamefont {McHugh}, \citenamefont
  {Chantrell},\ and\ \citenamefont {D'Amico}}]{marchetti2014}%
  \BibitemOpen
  \bibfield  {author} {\bibinfo {author} {\bibfnamefont {G.}~\bibnamefont
  {Marchetti}}, \bibinfo {author} {\bibfnamefont {M.}~\bibnamefont {Hodgson}},
  \bibinfo {author} {\bibfnamefont {J.}~\bibnamefont {McHugh}}, \bibinfo
  {author} {\bibfnamefont {R.}~\bibnamefont {Chantrell}}, \ and\ \bibinfo
  {author} {\bibfnamefont {I.}~\bibnamefont {D'Amico}},\ }\href@noop {}
  {\bibfield  {journal} {\bibinfo  {journal} {Materials}\ }\textbf {\bibinfo
  {volume} {7}},\ \bibinfo {pages} {2795} (\bibinfo {year} {2014})}\BibitemShut
  {NoStop}%
\bibitem [{\citenamefont {Friedel}(1958)}]{friedel1958}%
  \BibitemOpen
  \bibfield  {author} {\bibinfo {author} {\bibfnamefont {J.}~\bibnamefont
  {Friedel}},\ }\href {\doibase 10.1007/BF02751483} {\bibfield  {journal}
  {\bibinfo  {journal} {Il Nuovo Cimento}\ }\textbf {\bibinfo {volume} {7}},\
  \bibinfo {pages} {287} (\bibinfo {year} {1958})}\BibitemShut {NoStop}%
\bibitem [{\citenamefont {Mahan}(2000)}]{mahan2000}%
  \BibitemOpen
  \bibfield  {author} {\bibinfo {author} {\bibfnamefont {G.~D.}\ \bibnamefont
  {Mahan}},\ }\href@noop {} {\emph {\bibinfo {title} {{Many-Particle
  Physics}}}}\ (\bibinfo  {publisher} {Kluwer Academic},\ \bibinfo {address}
  {New York},\ \bibinfo {year} {2000})\BibitemShut {NoStop}%
\bibitem [{\citenamefont {Stern}(1967)}]{stern1967}%
  \BibitemOpen
  \bibfield  {author} {\bibinfo {author} {\bibfnamefont {F.}~\bibnamefont
  {Stern}},\ }\href {\doibase 10.1103/PhysRev.158.697} {\bibfield  {journal}
  {\bibinfo  {journal} {Phys. Rev.}\ }\textbf {\bibinfo {volume} {158}},\
  \bibinfo {pages} {697} (\bibinfo {year} {1967})}\BibitemShut {NoStop}%
\bibitem [{\citenamefont {Patterson}\ and\ \citenamefont
  {Lehoczky}(1989)}]{patterson1989}%
  \BibitemOpen
  \bibfield  {author} {\bibinfo {author} {\bibfnamefont {J.~D.}\ \bibnamefont
  {Patterson}}\ and\ \bibinfo {author} {\bibfnamefont {S.~L.}\ \bibnamefont
  {Lehoczky}},\ }\href@noop {} {\bibfield  {journal} {\bibinfo  {journal}
  {Physics Letters A}\ }\textbf {\bibinfo {volume} {137}},\ \bibinfo {pages}
  {137 } (\bibinfo {year} {1989})}\BibitemShut {NoStop}%
\bibitem [{\citenamefont {Landau}\ and\ \citenamefont
  {Lifshitz}(1977)}]{landau1977}%
  \BibitemOpen
  \bibfield  {author} {\bibinfo {author} {\bibfnamefont {L.~D.}\ \bibnamefont
  {Landau}}\ and\ \bibinfo {author} {\bibfnamefont {E.~M.}\ \bibnamefont
  {Lifshitz}},\ }\href@noop {} {\emph {\bibinfo {title} {{Quantum Mechanics
  (Third Edition, Revised and Enlarged)}}}}\ (\bibinfo  {publisher}
  {Pergamon},\ \bibinfo {address} {Oxford},\ \bibinfo {year}
  {1977})\BibitemShut {NoStop}%
\bibitem [{\citenamefont {Pohv}\ \emph {et~al.}(2002)\citenamefont {Pohv},
  \citenamefont {Rith}, \citenamefont {Scholz},\ and\ \citenamefont
  {Zetsche}}]{povh2002}%
  \BibitemOpen
  \bibfield  {author} {\bibinfo {author} {\bibfnamefont {B.}~\bibnamefont
  {Pohv}}, \bibinfo {author} {\bibfnamefont {K.}~\bibnamefont {Rith}}, \bibinfo
  {author} {\bibfnamefont {C.}~\bibnamefont {Scholz}}, \ and\ \bibinfo {author}
  {\bibfnamefont {F.}~\bibnamefont {Zetsche}},\ }\href@noop {} {\emph {\bibinfo
  {title} {{Particles and Nuclei}}}},\ \bibinfo {edition} {1st}\ ed.\ (\bibinfo
   {publisher} {Springer},\ \bibinfo {address} {Berlin},\ \bibinfo {year}
  {2002})\BibitemShut {NoStop}%
\end{thebibliography}%

\bibliographystyle{apsrev4-1}

\end{document}